\documentclass[aps,prc,twocolumn,letterpaper,superscriptaddress,nofootinbib,showpacs,floatfix,10pt]{revtex4-1}
\usepackage[utf8]{inputenc}
\usepackage{graphicx}
\usepackage{comment}
\usepackage{setspace}
\usepackage{amsmath}
\usepackage{amssymb}
\usepackage[textwidth=16cm,textheight=22cm]{geometry}
\usepackage{color}
\usepackage{indentfirst}
\usepackage{xspace}
\usepackage{hyperref}
\usepackage{verbatim}
\usepackage{epstopdf}
\usepackage{graphicx}
\usepackage{longtable}
\usepackage{lineno}
\usepackage{subfigure}
\usepackage{parskip}
\usepackage{xcolor}

\newcommand{\pt}   {\ensuremath{p_\mathrm{T}}\xspace}

\newcommand{\pythia} {\textsc{pythia8}\xspace}

\begin{document}

\title{Jet-associated Balance Functions of Charged and Identified Hadrons in pp Collisions at $\sqrt{s}=13.6$ TeV using PYTHIA8}

\author{Subash Chandra Behera}
\email{subash.chandra.behera@cern.ch}
\affiliation{
INFN--Sezione di Roma, Piazzale Aldo Moro, 2 - 00185 Roma RM, Italy}
\author{Arvind Khuntia}
\email{arvind.khuntia@cern.ch}
\affiliation{INFN--Sezione di Bologna, via Irnerio 46, 40126 Bologna BO, Italy}

\begin{abstract}
We present a study of charge balance functions inside jets in proton–proton collisions at $\sqrt{s}=13.6$ TeV using the PYTHIA8 event generator. The balance function is a differential observable of opposite-charge correlations, which is calculated in the jet frame for inclusive charged hadrons and the identified $\pi$, $K$, and $p$. The results show a clear narrowing of the balancing width with increasing jet charged multiplicity, indicating that particle production becomes more localized in momentum space in high-multiplicity jets. This trend resembles features attributed to collective expansion in heavy-ion collisions. The species dependence highlights sensitivity to the redistribution of strangeness and baryon number during string fragmentation and color reconnection. The new CR tune yields a little broader proton balance-function width in $\Delta\phi^{*}$ than CP5, hinting at enhanced baryon-production dynamics, whereas meson widths differ only mildly. These comparisons suggest that multiparton interactions and color reconnection contribute to the observed trends, potentially generating collective like features inside jets, especially in high multiplicity jets, via nontrivial color dynamics alongside standard fragmentation. Taken together, the results establish identified hadron balance functions in high multiplicity jets as a sensitive probe of hadronization and provide new constraints for models of small system collectivity.
\end{abstract}


\maketitle

\section{Introduction}

The study of hadronization remains one of the most challenging aspects of quantum chromodynamics (QCD). While perturbative techniques successfully describe the early stages of a parton shower, the transition to color-neutral hadrons involves nonperturbative processes that are not yet fully understood. Traditional models, such as string or cluster fragmentation, reproduce many features of the data. Jets, as collimated sprays of hadrons originating from high-energy partons, offer a controlled environment to investigate hadronization under extreme local conditions. In particular, jets with high charged-particle multiplicities or those produced in high-energy proton–proton (pp) collisions can generate localized regions of high energy density approaching or even exceeding that of central heavy-ion collisions on small spatial scales and short timescales \cite{ALICE:2020fuk}. Within such dense and dynamic environments, overlapping color strings or partons may undergo final-state interactions that modify hadron production, leading to effects such as strangeness enhancement, radial flow–like signatures, or changes in baryon-to-meson ratios.  Recent studies have proposed that a small, but strongly interacting partonic system may arise even from the fragmentation of a single high-energy parton moving through the quantum chromodynamic vacuum. In this picture, the strong color field of a propagating parton can generate additional partons that rescatter and expand collectively \cite{Baty:2021ugw}. This may develop collective dynamics , leading to azimuthal anisotropy or even radial flow signatures. This is akin to Monte Carlo study that indicate that \pythia produces radial-flow–like behavior in dense partonic environments arising from multiparton interactions (MPI) and string interactions via color reconnection (CR) \cite{OrtizVelasquez:pythaiaFlowLike}.\\

Motivated by this concept, the CMS experiment recently performed a study of angular correlations between charged particles within individual jets produced in pp collisions at $\sqrt{s}$ = 13 TeV \cite{CMS:2025:jetFlow}. By defining a new coordinate system centered on the jet axis, two-particle correlation functions in terms of relative azimuthal angle and pseudorapidity are measured. The results revealed that the strength of azimuthal anisotropy, quantified by the second harmonic coefficient, increases with the number of charged particles within a jet. This upward trend becomes particularly pronounced in jets with more than eighty constituents. While conventional models such as \pythia can reproduce the decrease of anisotropy at low jet multiplicity, they fail to describe the observed enhancement at higher multiplicities \cite{CMS:2025:jetFlow}. This discrepancy suggests the emergence of non-trivial final state effects within jets, possibly related to local parton density, quantum interference, or string dynamics. These findings point toward the possibility that jets with large multiplicities may serve as a new probe for exploring collective phenomena in small systems.\\

Charge correlations are a powerful tool to study particle production mechanisms in high-energy collisions~\cite{Bass1, cmsbf, alicebfpbpbpb, Pruneau:2019baa, ALICE:pruneau, scottpratt}.
The balance function is a differential observable that quantifies the correlation between opposite-sign charged particle pairs as a function of their relative momentum or angular separation~\cite{Pratt2012dz, Pruneau:new, Pruneau:2019baa, STAR:2010plm, Behera:2025ycg, Behera:2025ymi}. Initially introduced to study charge conservation and hadronization time scales in heavy-ion systems. Balance functions are increasingly being applied to more localized environments, such as jets, where they offer insight into parton fragmentation. Jets, originating from high-\pt partons produced in hard QCD scatterings, serve as precise probes of the hadronization process. Inside a jet cone, balance functions reveal how quark-antiquark pairs are distributed during fragmentation, how flavor and mass affect charge correlations, and whether local charge conservation is preserved within the collimated partonic shower~\cite{Baty:2021ugw, cmswhite, CMS:2023iam}.\\

While earlier studies have focused mainly on angular correlations and flow coefficients, the balance function provides a more differential way to probe how opposite charge pairs are distributed inside jets. Its sensitivity to the relative separation of these pairs offers insight into the time and mechanism of hadronization as well as the role of final state interactions~\cite{cmsbf, Bialas:2, alicethree}. Pions are the most abundant light mesons and trace the bulk of fragmentation. Kaons carry strangeness and give information on strange quark production and transport. Protons reflect baryon number conservation and possible differences between baryons and mesons in the hadronization process~\cite{scottpratt, ALICE:pruneau, Behera:2025ycg}. Studying the balance function separately for pions, kaons and protons as a function of jet multiplicity allows one to explore whether the narrowing of correlations with increasing multiplicity depends on particle species. The inclusion of particle identification therefore, adds an important dimension to jet-based correlation studies and may help to distinguish between different mechanisms of QCD in small systems.\\

In this work, the balance function is measured for different charged hadron species, including pions, kaons, and protons, as well as unidentified charged hadrons, as a function of jet charge multiplicity. Jets are reconstructed using the anti-$k_{T}$ algorithm of FastJet framework~\cite{FastJetMan} with a resolution parameter $R=0.8$, and the analysis is performed in the jet frame within a pseudorapidity coverage of $|\eta^{*}| < 2.4$. High-$p_{\mathrm{T}}$ jets are selected since they originate from hard parton scatterings, providing a well-defined axis and a cleaner environment that minimizes underlying-event contributions. With this setup, we investigate how local charge conservation manifests at the level of jet fragmentation and whether species-dependent effects emerge in high-multiplicity jets.\\

This paper is organized as follows. Section~\ref{anaproedure} discusses the analysis method to construct the balance functions with respect to the jet frame. Section~\ref{modeldes} demonstrates the \pythia model calculations and tune settings used for this study. Section~\ref{results} presents the results of the balance functions of $\pi$, K and $\it{p}$ as well as charged hadrons in terms of $\Delta\eta^{*}$ and $\Delta\phi^{*}$, and their widths as a function of jet charged-particle multiplicity $(N_\mathrm{ch}^{j})$. Section~\ref{summary} summarizes the findings of this work.

\section{Analysis Methodology}
\label{anaproedure}

In this study, we examine two-particle angular correlations in the reference frame of reconstructed jets. The analysis methodology is conceptually similar to that used in Ref.~\cite{Baty:2021ugw, Vertesi:2024fwl, CMS:2023iam}, where all particle momentum vectors are redefined relative to the jet axis. The two-dimensional correlation function is then calculated as a function of the relative azimuthal angle $\Delta\phi^{*}$ and pseudorapidity $\Delta\eta^{*}$ between particle pairs within the jet. The correlation functions are calculated from the ratio of signal and mixed-event distributions, following the standard approaches in previous studies~\cite{Baty:2021ugw, CMS:2023iam, cmsbf, cmsppridge, cmspbpbflow, hin18008, CMSPP, Behera:2025ycg, Behera:2025ymi}. The \text{signal distribution}, denoted as \( S(\Delta \eta^{*}, \Delta \phi^{*}) \), is constructed by pairing particles within the same event and is defined as
\begin{equation} \label{eqn_sig}
S(\Delta \eta^{*}, \Delta \phi^{*}) = \frac{1}{N_\text{ch}^\text{trg}} \frac{d^{2}N^\text{same}}{d\Delta \eta^{*} \, d\Delta \phi^{*}},
\end{equation}
where \( N_\text{ch}^\text{trg}\) represents the number of trigger particles in the \( j_{\mathrm{T}} \) interval of $0.3 < j_\mathrm{T} < 3.0$ GeV and $|\eta_{jet}^{*}| \leq 1.6$. \( N^\text{same} \) is the number of particle pairs binned in \( \Delta \eta^{*} \) and \( \Delta\phi^{*} \). A conventional event-mixing technique is used to generate the mixed-event distribution, $M(\Delta \eta^{*}, \Delta\phi^{*})$.  In this approach, trigger particles from a given event are combined with associated particles chosen from a set of ten randomly selected events. The mixed event distribution is defined as 
\begin{equation} \label{eqn_mix} 
M(\Delta \eta^{*},\Delta \phi^{*}) = \frac{1}{N_\text{ch}^\text{trg}}\frac{d^{2}N^\text{mix}}{{d\Delta \eta^{*}} \ {d\Delta \phi^{*}}},
\end{equation}
where the number of mixed event pairs for a given $\Delta \eta^{*}$ and $\Delta\phi^{*}$ bin is denoted by $N_\text{mix}$. The mixed event distribution corrects for acceptance effects based on by the detector's finite $\eta^{*}$ range and provides an analogy to measure correlations. The angular correlation functions in two dimensions are defined as 
\begin{equation} \label{eqn_2pc}
\frac{1}{N_\text{ch}^\text{trg}}\frac{d^{2}N^\text{pair}}{{d\Delta \eta^{*}} \ {d\Delta \phi^{*}}} = M(0,0) \frac{S(\Delta \eta^{*}, \Delta\phi^{*})}{M(\Delta \eta^{*}, \Delta\phi^{*})}.
\end{equation}
In this formulation, the effects of pair acceptance are largely accounted for by the ratio $\frac{M(0,0)}{M(\Delta y^{*}, \Delta \phi^{*})}$, where $M(0,0)$ denotes the mixed-event yield of particle pairs emitted in nearly the same direction. This region corresponds to the highest detection efficiency for pairs~\cite{cmspbpbflow, cmsbf, cmsppridge}. The physical interpretation of the correlations depends on the charge combination of the particle pair. Unlike-sign pairs are primarily sensitive to local charge conservation and thus sensitive to charge-balancing effects. They also have contributions from resonance decays (e.g. $\rho^0 \to \pi^+\pi^-$, $K^{*0} \to K^+\pi^-$), where decay kinematics naturally produce correlated opposite-charge partners~\cite{STAR:2010plm, aliceoldpbpb, cmsbf, Bass1, Pratt:2003gh, CMS:2013jlh}. In addition, the Coulomb attraction between oppositely charged particles can enhance correlations at small relative momentum. Like-sign pairs can probe other important mechanisms, such as Bose–Einstein quantum interference (Hanbury-Brown-Twiss correlations) leads to a near-side enhancement for identical bosons (e.g. $\pi^+\pi^+$ or $\pi^-\pi^-$)~\cite{Pratt:fem, alice_fem, hbt, Jeon:2001ue}. Additionally, Coulomb repulsion between like-charged particles modifies the correlation structure.\\

The balance function represents the conditional probability of observing a particle of opposite charge at a given angular separation from a reference particle. Its general form can be written as
\begin{equation} \label{eqn_balfun}
B^{\alpha \beta}=\frac{1}{2}[C_{2}^{\alpha^{+} \beta^{-}} + C_{2}^{\alpha^{-} \beta^{+}} - C_{2}^{\alpha^{-} \beta^{-}} - C_{2}^{\alpha^{+} \beta^{+}}],
\end{equation}

where $\alpha$ and $\beta$ denote the hadron species. The $+$ and $-$ signs indicate the electric charge of the particles. Each term $C_{2}^{\alpha^q\beta^{q'}}$ represents a two-particle correlation function for a particular charge combination. The unlike-sign terms capture the correlations directly linked to local charge conservation, while subtraction of the like-sign contributions reduces charge-independent backgrounds, thereby isolating the balancing signal. From the balance functions for different identified hadron species, one can separately investigate the conservation of various quantum numbers. For example, $B^{\pi\pi}$ probes the balancing of light quark charges, $B^{KK}$ provides access to correlations associated with strangeness production and hadronization, and $B^{pp}$ reflects baryon number conservation and transport mechanisms.\\

To characterize the extent of charge-dependent correlations, the width of the balance function in $\Delta\eta^{*}$ or $\Delta\phi^{*}$ is determined~\cite{alicebfpbpbpb, alicep2r2pp, Behera:2025ycg, Behera:2025ymi}. The root-mean-square (RMS) width is obtained from the one-dimensional projection of $B$, assuming that the distribution is symmetric and centered around zero.
\begin{equation}
\sigma({\Delta \eta^{*}}) = \sqrt{ \frac{\sum_i B(\Delta \eta_{i}^{*}) \, (\Delta \eta_{i}^{*})^{2}}{\sum_i B(\Delta \eta_{i}^{*})} } ,
\label{eq:bf_width}
\end{equation}
where $B(\Delta \eta_{i}^{*})$ is the balance function value in the bin centered at $\Delta \eta_{i}^{*}$, and the summation extends over all bins of the measured distribution. A similar expression is used to estimate the width in $\Delta\phi^{*}$. In this analysis, the widths of $B$ are determined from one-dimensional projections within $|\Delta \eta^{*}| \leq 2.0$ and $|\Delta\phi^{*}| \leq 1.5$. For studies at higher transverse momentum, the evaluation is restricted to $|\Delta \eta^{*}|, |\Delta\phi^{*}| \leq 1.5$ to optimize statistical precision while retaining sensitivity to the underlying correlation structures.
\begin{figure*}[!htb]  
    \centering
    \subfigure[]{
        \includegraphics[width=0.9\textwidth]{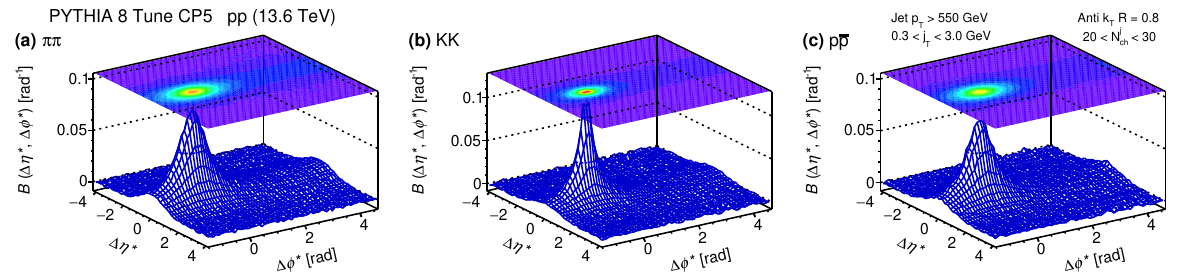}
    }
    \subfigure[]{
        \includegraphics[width=0.9\textwidth]{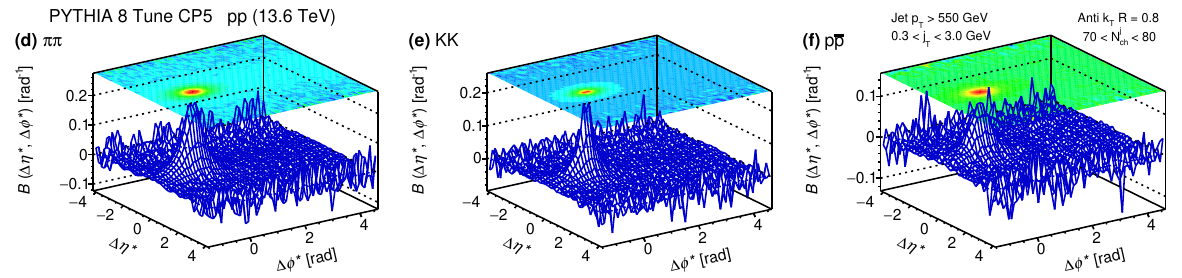}
    }
 \caption{Two-dimensional balance functions from tune CP5 in \pythia simulations for $\pi$, $K$, and $p$ in $pp$ collisions at $\sqrt{s} = 13.6$ TeV within individual jets. The top panel corresponds to $B(\Delta\eta^{*}, \Delta\phi^{*})$ in $20< N_\mathrm{ch}^{j} < 30$ intervals, and the bottom panel is for the $70<  N_\mathrm{ch}^{j}< 80$ range. The left column (a, d) presents the $B(\Delta\eta^{*}, \Delta\phi^{*})$ for pions, the middle column (b, e) is for kaons, and the right column (c, f) is for protons. Both the trigger and associated particles are considered in  $0.3 < j_\mathrm{T} < 3.0$ GeV.}
    \label{fig:cbf_pythia}
\end{figure*}

\begin{figure*}[!htb]  
    \centering
    \subfigure[]{
        \includegraphics[width=0.9\textwidth]{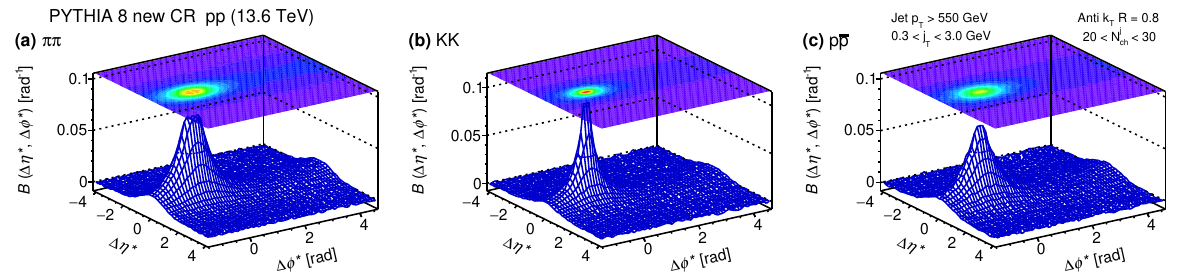}
    }
    \subfigure[]{
        \includegraphics[width=0.9\textwidth]{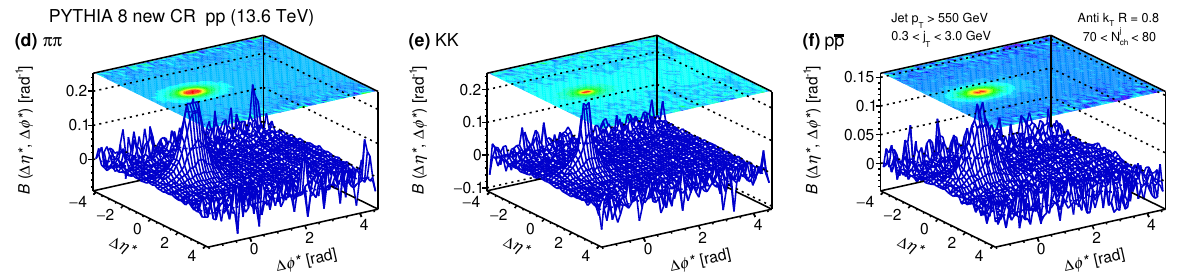}
    }
 \caption{Two-dimensional balance functions from new CR in \pythia simulations for $\pi$, $K$, and $p$ in $pp$ collisions at $\sqrt{s} = 13.6$ TeV within individual jets. The top panel corresponds to $B(\Delta\eta^{*}, \Delta\phi^{*})$ in $20< N_\mathrm{ch}^{j} < 30$ intervals, and the bottom panel is for the $70< N_\mathrm{ch}^{j} < 80$ range. The left column (a, d) presents the $B(\Delta\eta^{*}, \Delta\phi^{*})$ for pions, the middle column (b, e) is for kaons, and the right column (c, f) is for protons. Both the trigger and associated particles are considered in  $0.3 < j_\mathrm{T} < 3.0$ GeV.}
    \label{fig:cbf_pythia_tuneM2}
\end{figure*}

\section{Model description}
\label{modeldes}
In this study, the Monte Carlo event generator \pythia is used to simulate pp collisions at $\sqrt{s}$ = 13.6 TeV, which employs the Lund string fragmentation model \cite{pythia8LundString}. Two different tunes are considered in order to investigate the sensitivity of the balance function to the modeling of soft-QCD processes: the CP5 tune and the recently developed color reconnection scheme. The CP5 tune is widely used and recommended for studying at the LHC energies. It is based on the NNPDF3.1 NNLO parton distribution functions. The parameters of CP5 are optimized using a wide range of data, including minimum-bias, underlying event, and multiparton interaction, ensuring a consistent description across different collisions. 

The new CR model implemented in \pythia introduces an alternative description of color reconnection, motivated by the need to improve the modeling of collective-like effects in small systems. The new CR model in \pythia \cite{Bierlich:newcr} allows reconnections that create junction topologies. Because junctions are closely tied to baryon production, the model naturally predicts a baryon enhancement and has been shown to reproduce $\Lambda$ yields simultaneously at LEP and the LHC. The new CR scheme effectively introduces two parameters: one that sets the overall strength of color reconnection and one that controls the baryon enhancement, both of which were tuned to 7 TeV LHC data.
 

\begin{figure*}[!ht]  
    \centering  
    \subfigure[]{
        \includegraphics[width=0.4\textwidth]{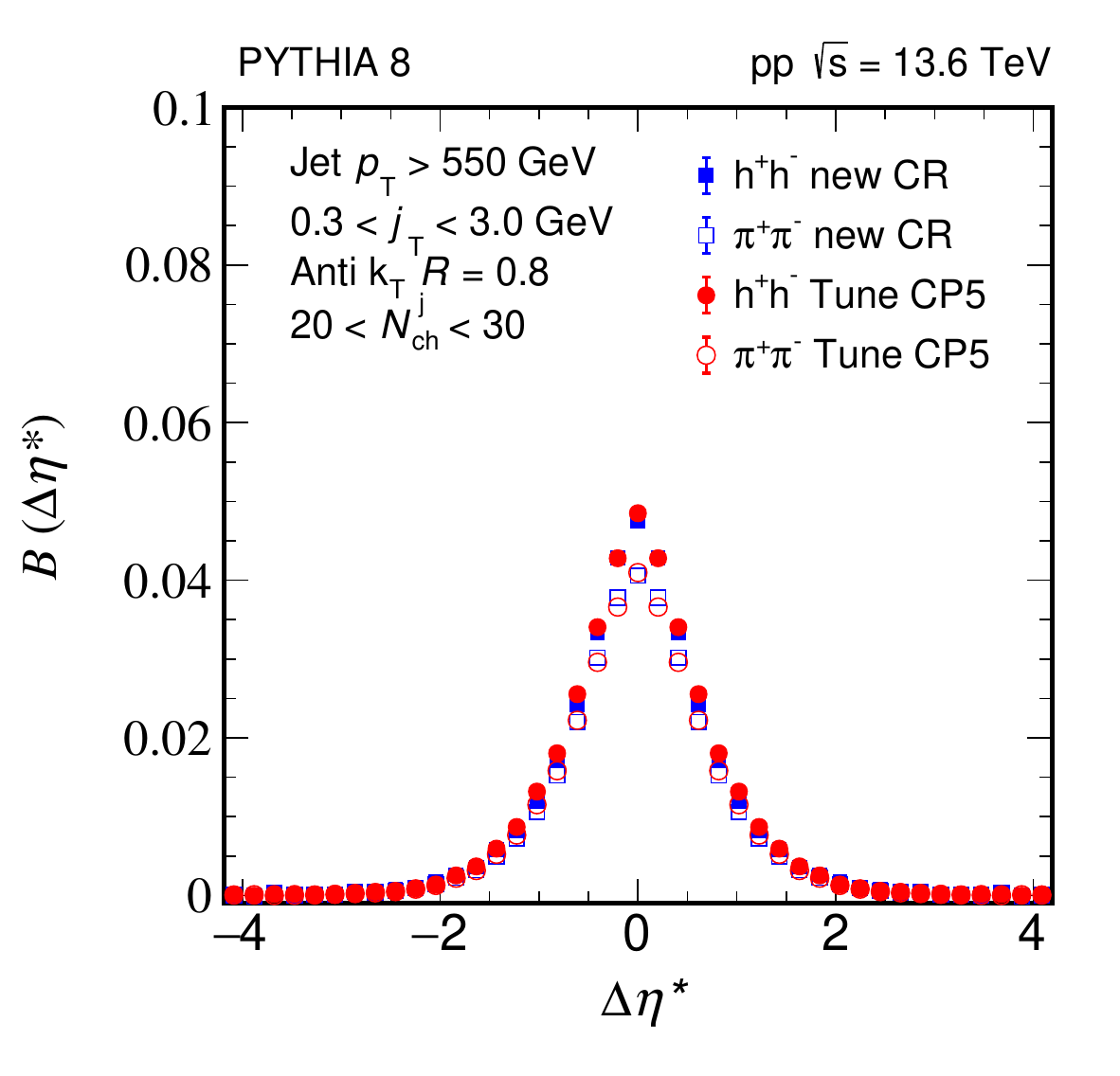}
    }
    \subfigure[]{
        \includegraphics[width=0.4\textwidth]{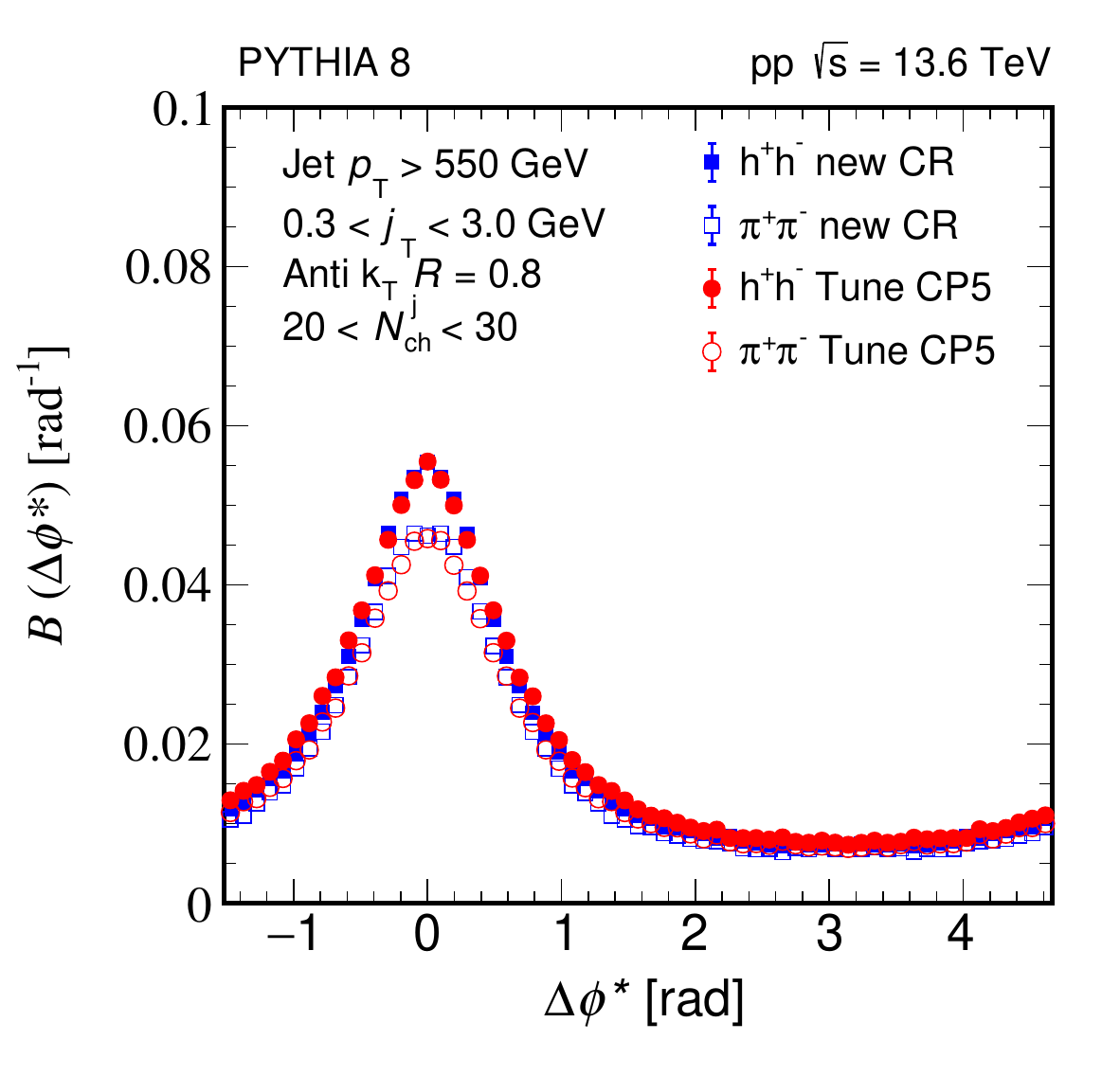}
    }
    \subfigure[]{
        \includegraphics[width=0.4\textwidth]{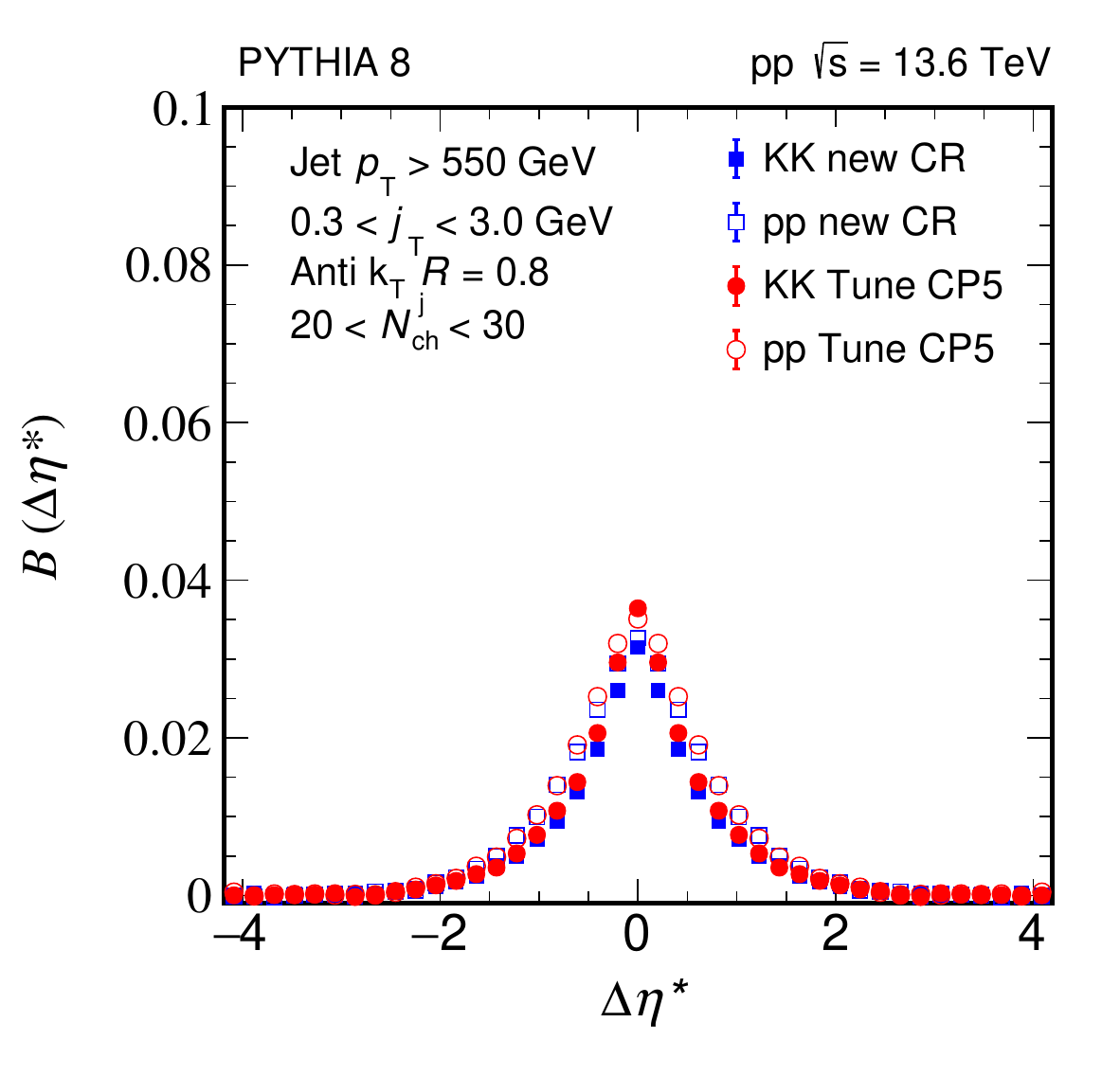}
    }
    \subfigure[]{
        \includegraphics[width=0.4\textwidth]{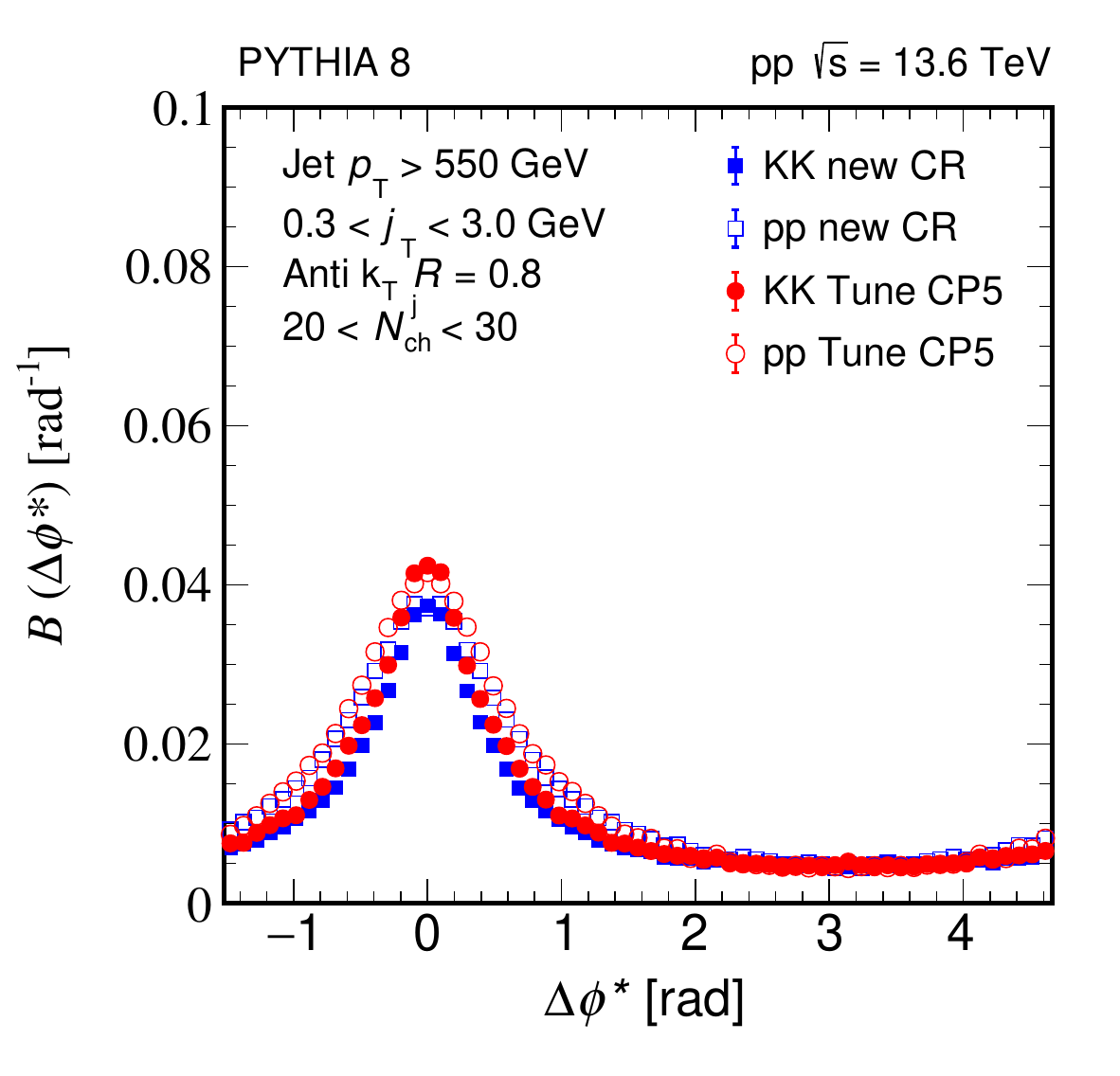}
    }

 \caption{One-dimensional $B$ projections along $\Delta\eta^{*}$ and $\Delta\phi^{*}$ from \pythia model calculations in $pp$ collisions at $\sqrt{s} =$ 13.6 TeV for different particle species. $\Delta\eta^{*}$ projections are taken in $|\Delta\phi|^{*} \leq \pi/2$ and $\Delta\phi^{*}$ projections are taken in the $|\Delta\eta^{*}| \leq 1.0$ range. The top row (a, b) shows $B(\Delta\eta^{*})$ and $B(\Delta\phi^{*})$ for pions and charged hadrons, while the bottom row (c, d) presents the corresponding projections for kaons and protons. The results are shown for events with $20 < N_{\mathrm{ch}}^{j} < 30$, and both the trigger and associated particles are considered in $0.3 < j_\mathrm{T} < 3.0~\mathrm{GeV}$.}
    \label{fig:model_plots_1d}
\end{figure*}

\section{Results and discussion}
\label{results}
\subsection{2D correlations}

Figure~\ref{fig:cbf_pythia}  and \ref{fig:cbf_pythia_tuneM2} present two-dimensional balance functions of $\pi$, $K$, and $p$ pairs within jets. The balance functions are evaluated for high jet-\pt, $p_{T}^{\mathrm{jet}} > 550$~GeV, with a charged particle multiplicity interval of $20 < N_{\mathrm{ch}}^{j} < 30$. The trigger and associated particles for both charged and identified hadrons are chosen in the range $0.3 < j_\mathrm{T} < 3.0$~GeV and $|\eta^{*}| < 2.4$.   The identified balance functions are shown for two different \pythia configurations, CP5 and new CR mode. In both cases, a pronounced near-side peak at $(\Delta\eta^{*}, \Delta\phi^{*}) \approx (0, 0)$ is observed for all particle species, reflecting the localized production of balancing charges inside jets. In the lower panel of Fig.~\ref{fig:cbf_pythia}, corresponding to higher multiplicity ($70 < N_{\mathrm{ch}}^{j} < 80$), the near-side peak becomes sharper and more collimated compared to the low-multiplicity case, indicating that balancing charges are increasingly localized in momentum space as the jet multiplicity grows. Figure~\ref{fig:cbf_pythia_tuneM2} shows similar distributions with the new CR configuration.


\subsection{One-dimensional projection and width estimation}

\begin{figure*}[!bth]  
    \centering
    \subfigure[]{
        \includegraphics[width=0.4\textwidth]{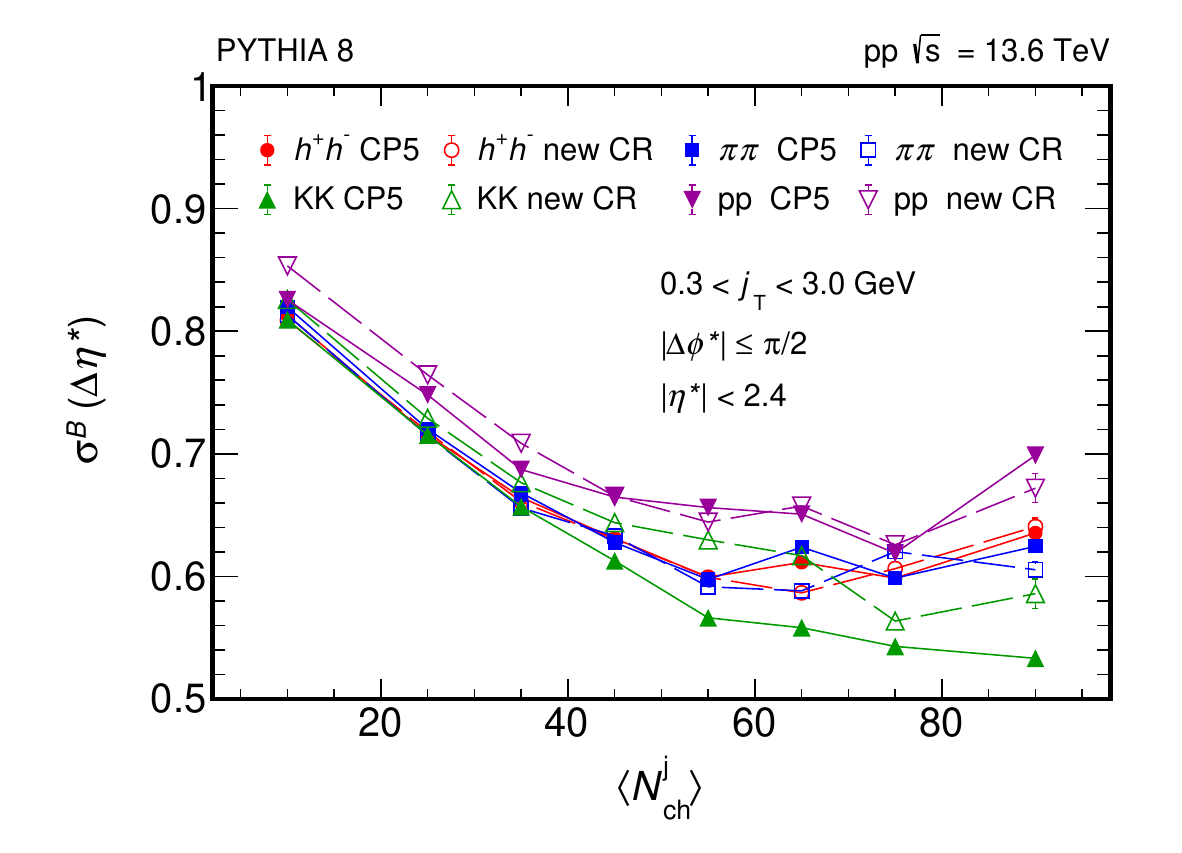}
    }
    \subfigure[]{
        \includegraphics[width=0.4\textwidth]{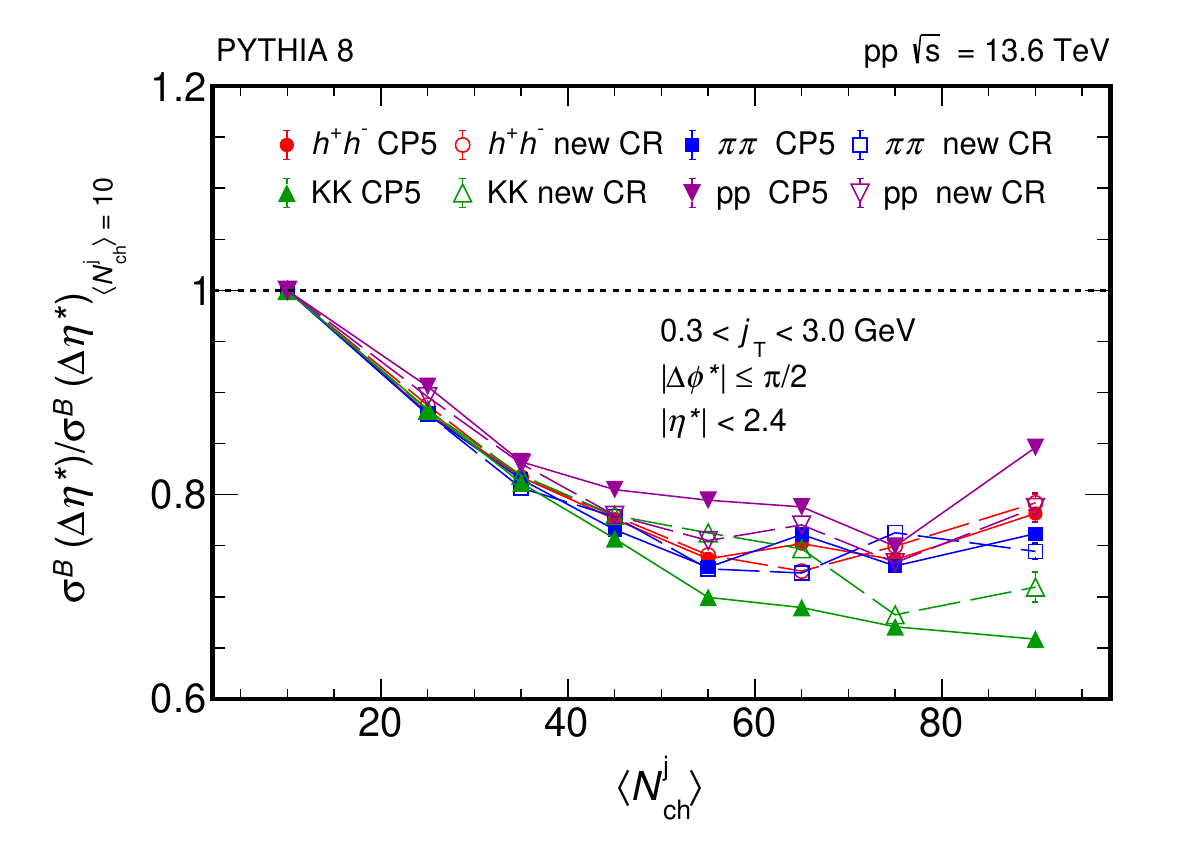}
    }
    \subfigure[]{
        \includegraphics[width=0.4\textwidth]{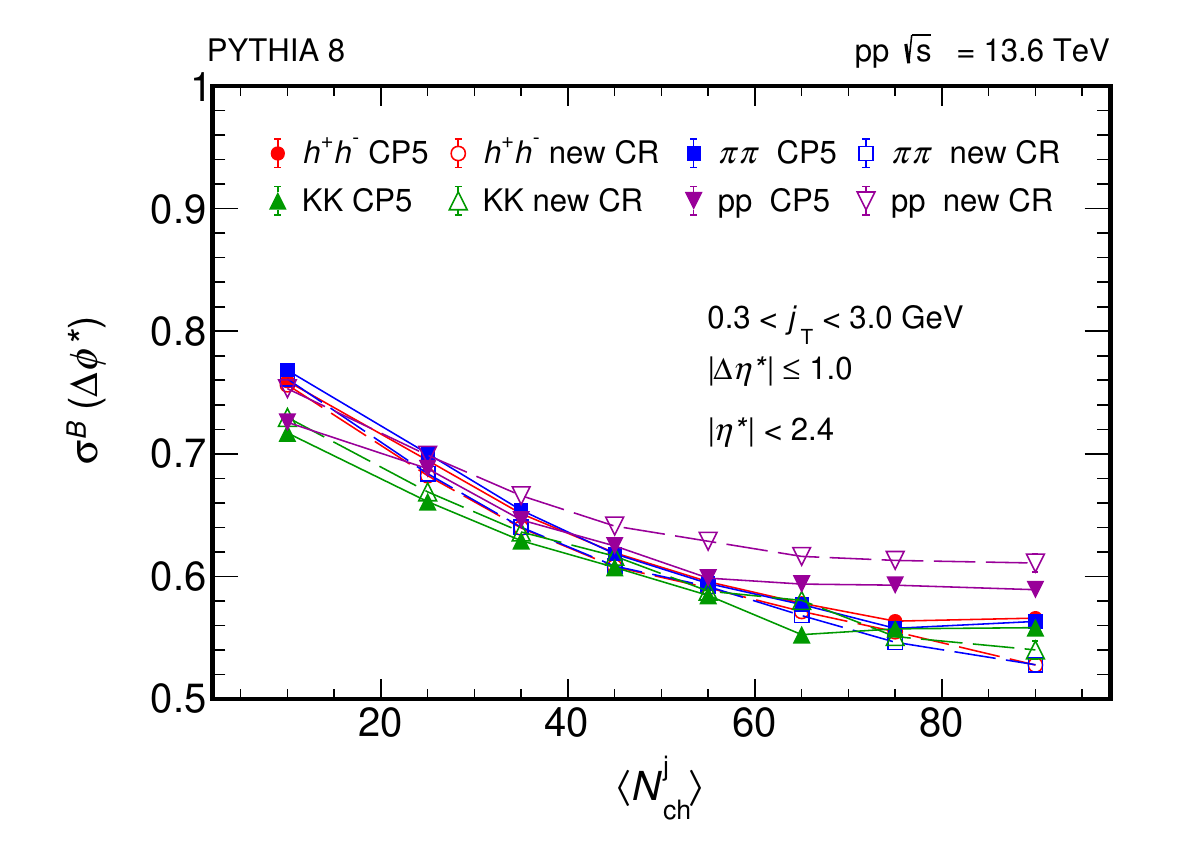}
    }
    \subfigure[]{
        \includegraphics[width=0.4\textwidth]{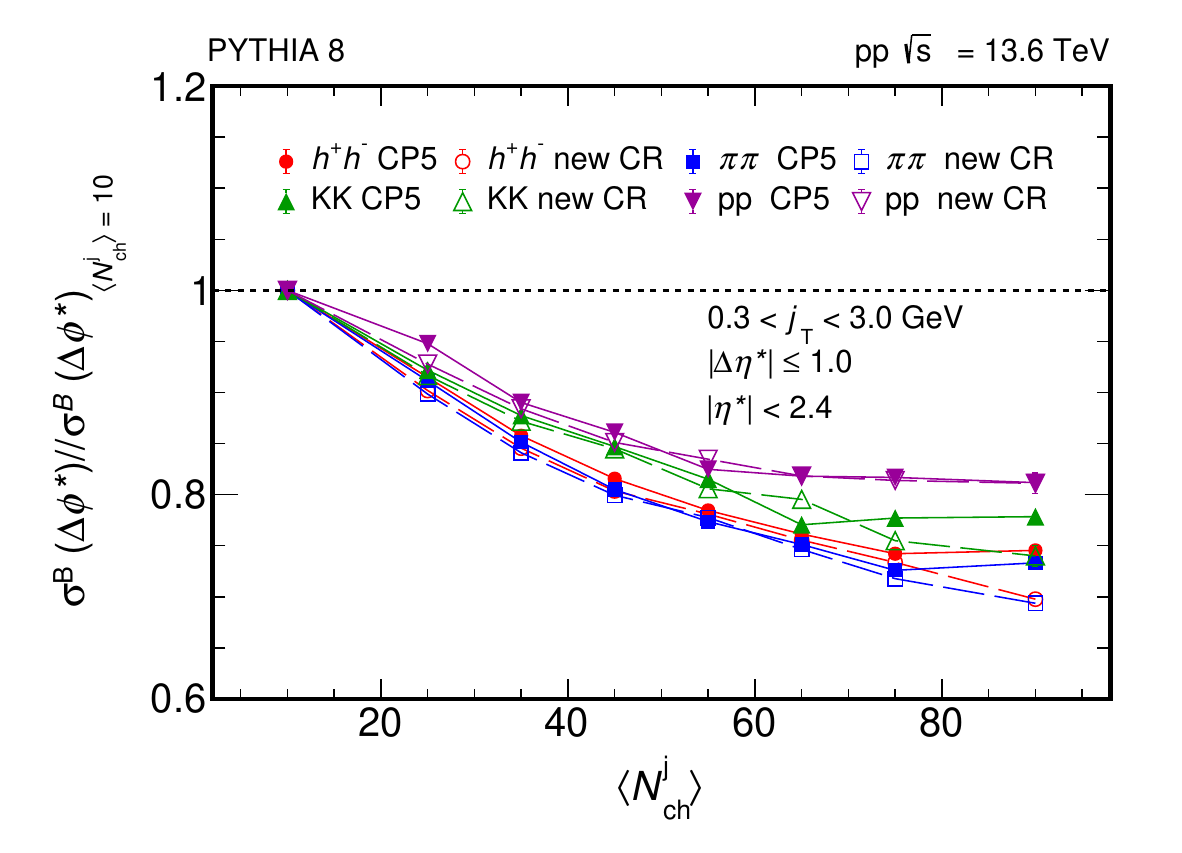}
    }
   
 \caption{The absolute (a, c) and the relative (b, d) width of $B$ projections along $\Delta\eta^{*}$ and $\Delta\phi^{*}$ from \pythia model calculations for charged-hadron, $\pi, K, p$, respectively. Both the trigger and associated particles are considered in $0.3 < j_\mathrm{T} < 3.0~\mathrm{GeV}$.}
    \label{fig:sigma_B_vs_Nch_ratio}
\end{figure*}

Figure~\ref{fig:model_plots_1d} presents the one-dimensional projections of  balance functions along $\Delta\eta^{*}$ and $\Delta\phi^{*}$ for different charged and identified species in $20 < N_\mathrm{ch}^{j} < 30$ multiplicity class. $\Delta\eta^{*}$ projections are made in $|\Delta\phi^{*}| < \pi/2$ and $\Delta\phi^{*}$ projections are calculated in $|\Delta\eta^{*} |< 1$. Panels~(a) and~(b) show, the projections of the balance function $B$ onto the $\Delta\eta^{*}$ and the $\Delta\phi^{*}$ for charged hadron pairs ($h^{\pm}$) and pion pairs ($\pi^{\pm}$). 
A pronounced near-side peak at $\Delta\eta^{*}$ $\approx 0$ and $\Delta\phi^{*}$ $\approx 0$ is observed for all the particle pairs, where charged hadron and pion balance functions show similar correlation strength. The same projections for kaon and proton pairs are shown in panels~(c) and~(d), where a narrower balance function distribution is observed in kaons as compared to protons. A comparison of balance functions constructed with and without resonance decays indicates only a mild dependence on the resonance contribution, with no significant impact on the extracted widths.

Figure~\ref{fig:sigma_B_vs_Nch_ratio} shows the multiplicity dependence of the jet-frame rms widths $\sigma_B$ of balance-function for $h^\pm$, $\pi^\pm$, $K^\pm$, and $p\bar p$ pairs in pp collisions at $\sqrt{s} = 13.6~\mathrm{TeV}$ using \pythia with both CP5 and new CR tunes. Both trigger and associated particles are selected with $0.3 < j_{\mathrm{T}} < 3.0~\mathrm{GeV}$. In the top panels, the pseudorapidity widths $\sigma_B(\Delta\eta^{*})$ decrease with increasing charged-particle multiplicity $\langle N^{j}_{\mathrm{ch}} \rangle$ for all hadron species, reflecting more localized charge balancing in denser jet environments. When normalized to the reference value at $\langle N^{j}_{\mathrm{ch}} \rangle = 10$, the relative drop at $\langle N_{\mathrm{ch}} \rangle \approx 70$ reaches about 25\% for pions, 30-35\% for kaons, and 18–20\% for protons. Charged hadrons follow a similar trend to pions, consistent with their dominance in the sample. The CP5 and new CR tunes yield comparable behavior in $\Delta\eta^{*}$, with only slight separation at high multiplicity. The bottom panels display the azimuthal widths $\sigma_B(\Delta\phi^{*})$ and their normalized ratios. A stronger tune dependence is observed, where all species show narrowing of width with multiplicity. At $\langle N^{j}_{\mathrm{ch}} \rangle \approx 70$, the normalized width decreases by up to 20-25\% for kaons, 20\% for protons, and 30\% for pions. This behavior is consistent with radial-flow–like azimuthal collimation driven by color-reconnection–induced boosts in the fragmenting string system. These results highlight the sensitivity of $\Delta\phi^{*}$ widths to hadron species and color-reconnection dynamics in the final state.

\subsection{\pt dependence study}
\begin{figure*}[!htb]  
    \centering
    \subfigure[]{
        \includegraphics[width=0.45\textwidth]{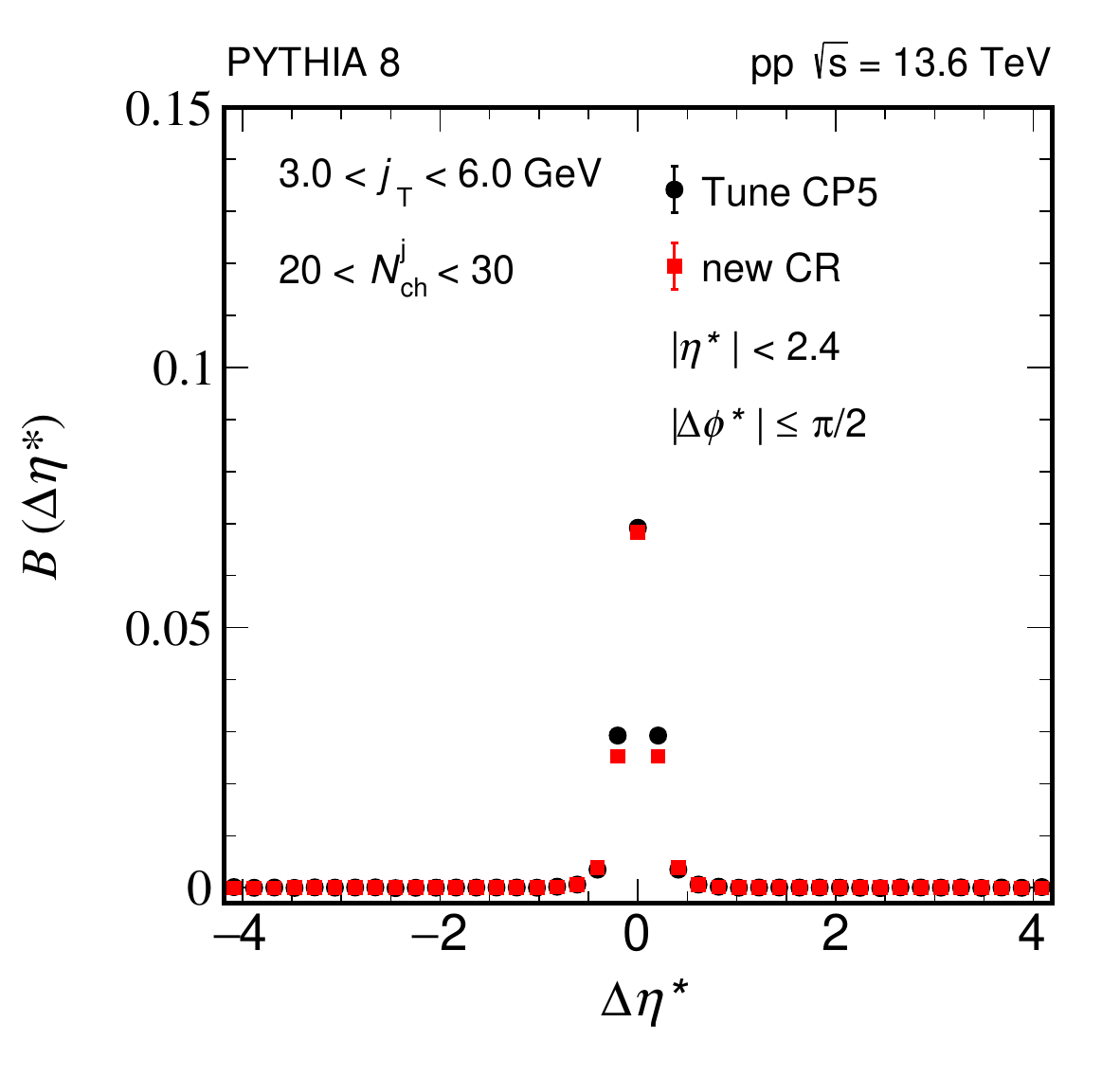}
    }
    \subfigure[]{
        \includegraphics[width=0.45\textwidth]{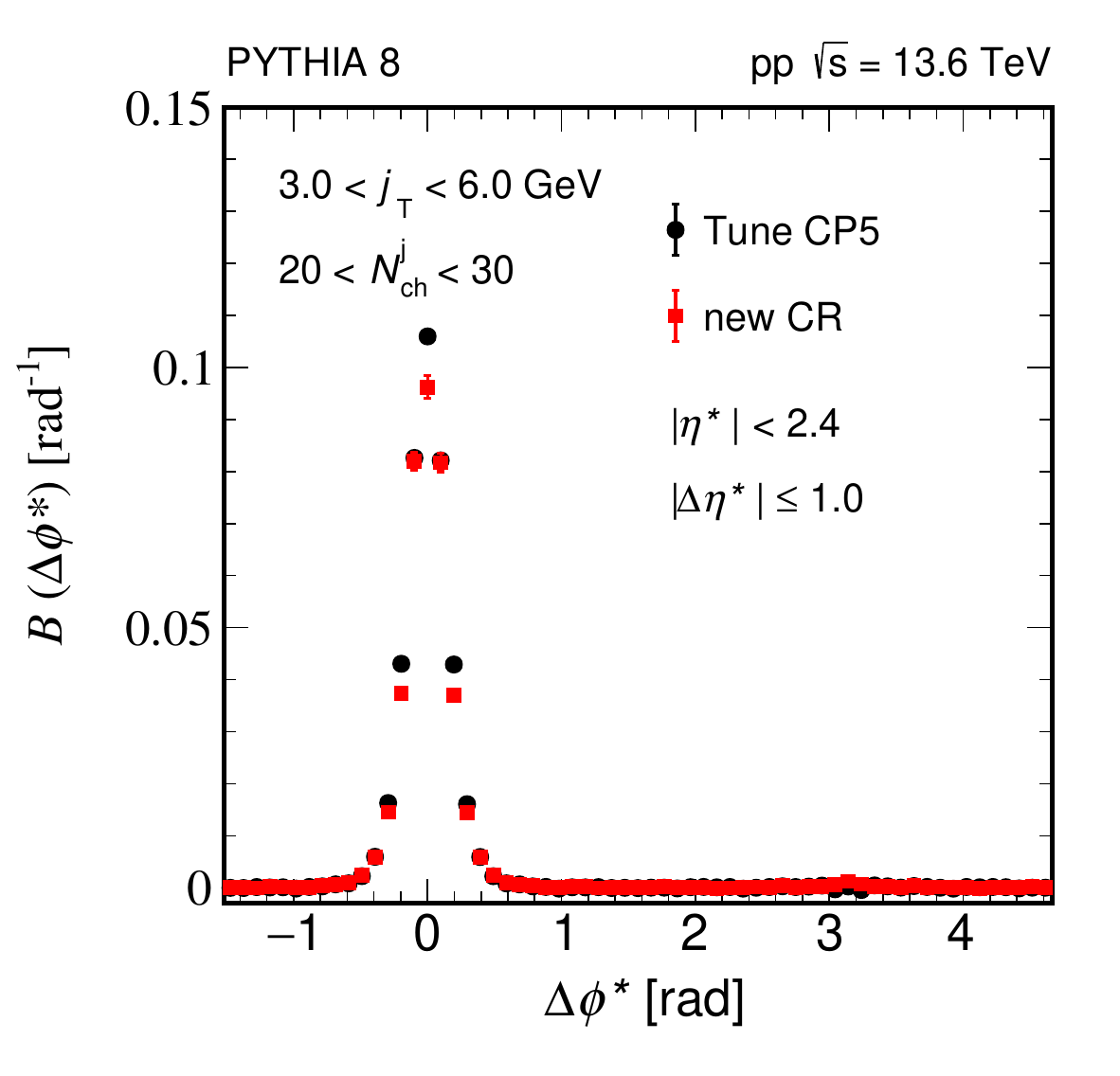}
    }
 \caption{One-dimensional projections of the balance functions for CP5 and the new CR tune from the \pythia model simulation of charged-hadrons in $pp$ collisions at $\sqrt{s} = $13.6 TeV. The left panel (a) shows the $B$ as a function of $\Delta\eta^{*}$, whereas the right panel (b) presents $B$ as a function of $\Delta\phi^{*}$. Both the trigger and associated particles are considered in $3.0 < j_\mathrm{T} < 6.0~\mathrm{GeV}$.}
    \label{fig:cbf_highpToneD_pythia}
\end{figure*}

The study of balance functions in the high transverse momentum regime inside jets provides a complementary perspective to traditional low‑$p_{\mathrm{T}}$ analyses. While at low-$p_{\mathrm{T}}$ the balance function is shaped primarily by soft QCD processes such as color reconnection and multi-parton interactions, at high-$p_{\mathrm{T}}$ both trigger and associated particles originate from the same jet and are dominated by hard partonic scatterings followed by jet fragmentation. Figure~\ref{fig:cbf_highpToneD_pythia} shows the one-dimensional projections of the balance function $B$ onto $\Delta\eta^{*}$ (left) and $\Delta\phi^{*}$ (right) for high-$p_{\mathrm{T}}$ hadron pairs inside jets, using \pythia with the CP5 and new CR tunes in the range $3.0<j_{\mathrm{T}}<6.0~\mathrm{GeV}$ with selections $|\eta^{*}|<2.4$. Compared to the low-$p_{\mathrm{T}}$ case, the balance functions are significantly narrower with sharply peaked structures near $(0,0)$, reflecting the localized nature of charge balancing from jet fragmentation. The similarity between CP5 and new CR indicates that color reconnection has only a minor impact in this high-$p_{\mathrm{T}}$ regime inside jets.\\

\begin{figure*}[!htb]  
    \centering
    \subfigure[]{
        \includegraphics[width=0.48\textwidth]{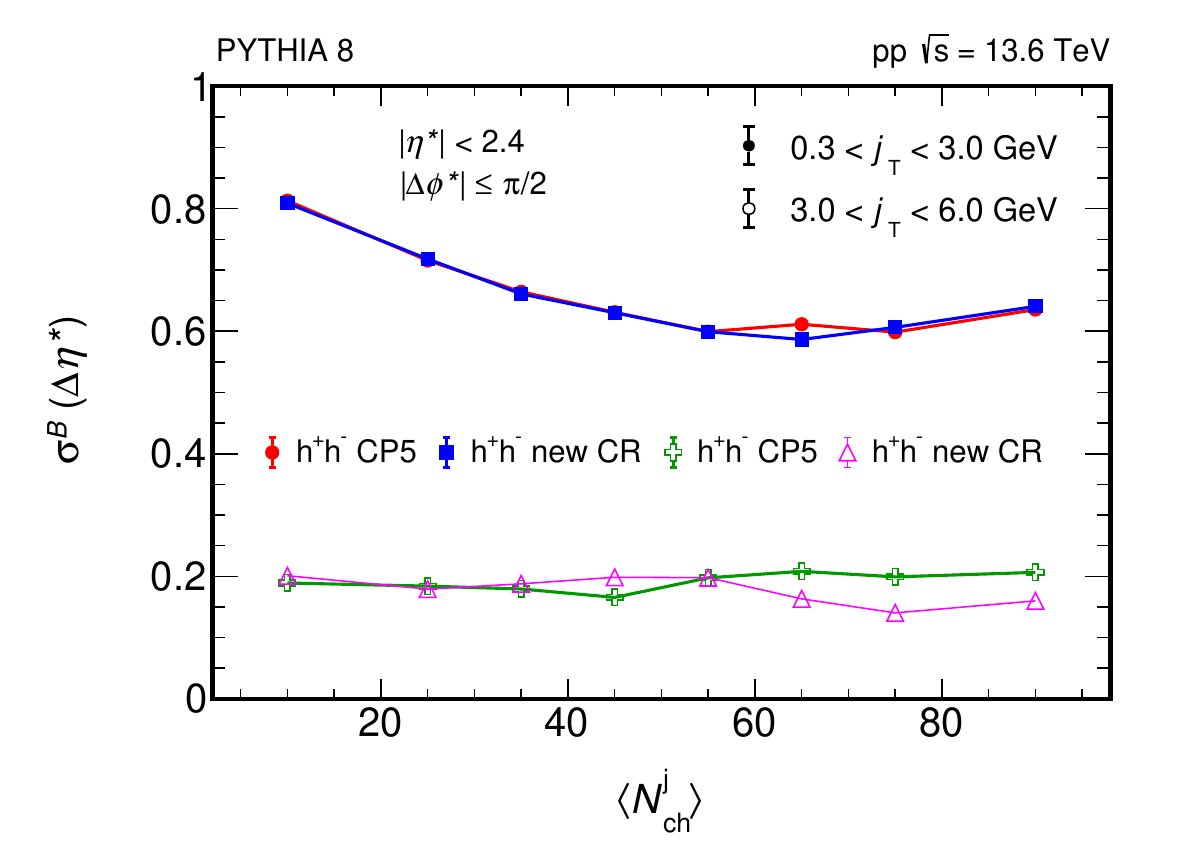}
    }
    \subfigure[]{
        \includegraphics[width=0.48\textwidth]{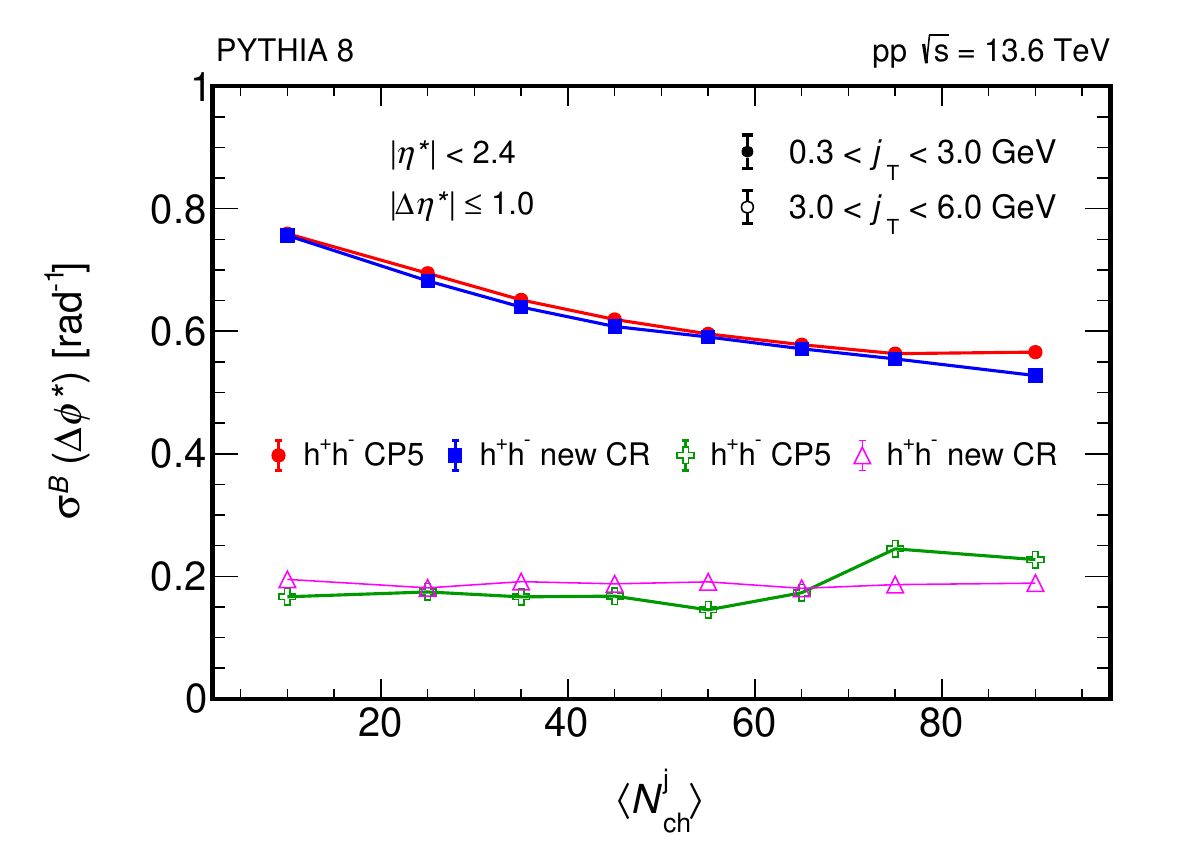}
    }
 \caption{The width comparison of the balance functions from the \pythia model simulation of charged-hadrons in pp collisions at $\sqrt{s} = $13.6 TeV. The left plot (a) shows the rms width in $\Delta\eta^{*}$ and the right plot (b) presents the rms width in $\Delta\phi^{*}$ for two different transverse momentum intervals as a function of $\langle N_{\mathrm{ch}}^{\text{j}} \rangle$.}
    \label{fig:sigmaB_high_low_pT}
\end{figure*}

Figure~\ref{fig:sigmaB_high_low_pT} presents the balance-function widths $\sigma^B(\Delta\eta^{*})$ and $\sigma^B(\Delta\phi^{*})$ as a function of jet-associated charged-particle multiplicity $\langle N_{\mathrm{ch}}^{\text{j}} \rangle$ for charged hadron species in pp collisions using the CP5 and new CR tunes. Two transverse momentum selections are shown: a low-$j_{\mathrm{T}}$ range ($0.3 < j_{\mathrm{T}} < 3.0~\mathrm{GeV}$) and a high-$j_{\mathrm{T}}$ range ($3.0 < j_{\mathrm{T}} < 6.0~\mathrm{GeV}$), both applied to trigger and associated particles within reconstructed jets. For the low-$j_{\mathrm{T}}$ selection, the widths in both $\Delta\eta^{*}$ and $\Delta\phi^{*}$ decrease with increasing jet multiplicity. This narrowing reflects stronger local charge conservation in more active jet environments and suggests the presence of final-state collective-like effects from color-reconnection induced boosts.

In contrast, the high-$j_{\mathrm{T}}$ selection shows an almost flat dependence of the balance-function widths on jet multiplicity. This indicates that in this regime, the charge-balancing partners originate from hard jet fragmentation and are largely decoupled from the overall jet activity or soft final-state effects. The similarity between the CP5 and new CR tunes confirms that color reconnection has limited influence at high $j_{\mathrm{T}}$.
\section{Summary}
\label{summary}
We present a comprehensive study of balance functions for charged and identified hadron pairs inside reconstructed jets in proton–proton collisions at $\sqrt{s} = 13.6$~TeV, using \pythia simulations with the CP5 and new color reconnection tunes. Jets are reconstructed using the anti-$k_{\mathrm{T}}$ algorithm with $R = 0.8$, and the analysis focused on jets with $p_{\mathrm{T}}^{\text{jet}} > 550$~GeV. Balance functions are evaluated in the jet frame for charged hadrons and identified pion, kaon, and proton pairs, as a function of jet-associated charged-particle multiplicity. We investigated two distinct $j_{\mathrm{T}}$ intervals for both trigger and associated particles. In the low-$j_{\mathrm{T}}$ region ($0.3 < j_{\mathrm{T}} < 3.0$ GeV), the balance-function widths in $\Delta\eta^{*}$ and $\Delta\phi^{*}$ exhibit a clear narrowing with increasing jet multiplicity. This behavior is consistent with enhanced local charge conservation and correlations that resemble collective behavior, possibly driven by color reconnection dynamics in high-multiplicity jets. The narrowing is more pronounced in the azimuthal direction, suggesting sensitivity to reconnection-induced radial flow–like effects. In contrast, for high-$j_{\mathrm{T}}$ pairs ($3.0 < j_{\mathrm{T}}$$ < 6.0$~GeV), the balance-function widths remain nearly constant across multiplicity, indicating that charge balancing is dominated by localized jet fragmentation and is largely unaffected by soft-QCD dynamics or global event activity.

Compared with earlier jet-frame correlation and thermodynamic-fragmentation studies, this work introduces the particle-identified charge-balance function as a new, differential probe of charge conservation inside jets. While previous analyses focused on long-range anisotropy or baryon-to-meson ratios, our study quantitatively connects the multiplicity-dependent narrowing of $\Delta\eta^{*}$ and $\Delta\phi^{*}$ widths to the interplay of color reconnection and hadronization dynamics. This provides a complementary and more flavor-sensitive perspective on collectivity in jets, establishing a clear phenomenological bridge between the recent CMS measurements and theoretical string-density models. Taken together, these results establish particle-identified balance functions inside jets as a novel and sensitive observable for exploring hadronization mechanisms, the role of strangeness and baryon number redistribution during string fragmentation and color reconnection, and the potential onset of collective behavior at small scales. These findings provide a well-characterized model baseline and motivate forthcoming measurements at LHC and RHIC experiments.
\begin{acknowledgements}
S.C. Behera and A. Khuntia acknowledge the support under INFN Postdoctoral fellowship.
\end{acknowledgements}

\bibliographystyle{utphys}   
\bibliography{reference}

\end{document}